\documentclass[aps,twocolumn,prl,tightenlines,floatfix,showpacs]{revtex4}
\usepackage{graphicx}
\usepackage[english]{babel}
\usepackage{amsmath}
\usepackage{amssymb}
\usepackage{times}
\usepackage{verbatim}

\newcommand{\phik}{\varphi_{\mathbf{k}}}

\begin{document}

\title{Quasiparticle Interference in Temperature Dependent-STM As a Probe
of Superconducting Coherence}
\author{Dan Wulin$^{1}$, Yan He$^{1}$, Chih-Chun Chien$^{1}$,
Dirk K. Morr$^{1,2}$ and K. Levin$^{1}$} \affiliation{$^1$James
Frank Institute and Department of Physics, University of Chicago,
Chicago, Illinois 60637 \\ $^2$ Department of Physics, University of
Illinois at Chicago, Chicago, Illinois 60607}

\date{\today}
\begin{abstract}
In this paper we explore the behavior of
the quasi-particle interference pattern (QPI) of scanning tunneling microscopy
as a function of temperature, $T$.
After insuring a minimal consistency with photoemission,
we find that
the QPI pattern
is profoundly sensitive to quasi-particle coherence and that it manifests
two energy gap scales.
The nearly dispersionless QPI pattern above $T_c$ is consistent
with data on moderately underdoped cuprates.
To illustrate the important two energy scale physics we present predictions
of the QPI--inferred
energy gaps as a function of $T$ for future experiments on
moderately underdoped cuprates.
\end{abstract}

\maketitle

Recently, attention in the field of high temperature
superconductivity has turned to characterizing the superconducting
phase in the underdoped regime. This phase necessarily differs
from that of a conventional $d$-wave superconductor because components
of the gap smoothly evolve
through $T_c$. This leads to
a normal state gap or pseudogap above $T_c$. Owing to
this fact, and unlike a BCS superconductor where the order
parameter and the excitation gap are identical, there are very few ways 
to probe directly something as fundamental as
the superconducting order parameter.
Within the
moderately underdoped samples, which we consider throughout this
paper, recent angle resolved photoemission spectroscopy (ARPES)
experiments have reported \cite{ShenNature,Kanigel}
%Shen2006,Kaminski,ANLPRL}
novel signatures of superconducting order. Fermi arcs around the $d$-wave
nodes above $T_c$ rapidly collapse \cite{Kanigel} at the transition
to form point nodes. It was argued that, within the superconducting
phase, the temperature dependence of the nodal gap has finally
provided \cite{ShenNature} ``a direct and unambiguous observation of
a single particle gap tied to the superconducting transition". A
complementary and equally valuable probe is scanning tunneling
microscopy (STM) and the related quasi-particle interference (QPI)
spectroscopy \cite{Gomes07,Seamus,YazdaniPRB,Yazdani}. While this
probe, like ARPES, is generally not phase sensitive, a controversy
has arisen as to whether these techniques can, as argued
experimentally \cite{Yazdani,Seamus}, or cannot. as summarized
theoretically \cite{YazdaniPRB,FranzQPI} 
%in pre-formed
%pair scenarios
, distinguish coherent
superconducting order from pseudogap behavior.

The objective of this Letter is to provide some resolution to this
controversy by studying the temperature evolution of the QPI
pattern observed in STM experiments from the superconducting
ground state into the pseudogap phase at $T \gtrsim T_c$. We employ
a \textit{microscopically derived} preformed pair theory
\cite{ourreview} that accounts \cite{ourarpespapers}
for all of the complex momentum and temperature
dependence of the ARPES spectral gap as described above.
The QPI pattern is obtained using
the Fourier transform of the local density of states (LDOS)
associated with a single impurity.
%Our
%theoretical model accounts \cite{ourarpespapers} for the
%complex momentum and temperature dependence of the ARPES spectral gap
%\cite{ShenNature,Shen2006,Kaminski,ANLPRL} as outlined above.
A central result of our study is that, once compatibility with
ARPES experiments is incorporated, the observation of the so-called
``octet model" QPI \cite{DHLEE2,Capriotti03} (called Bogoliubov QPI
or B-QPI) is a direct signature of \textit{coherent} superconducting
order: sets of consistent octet model vectors can only be found
for energies less than the superconducting order parameter energy
scale. B-QPI does not persist above $T_c$, where the order parameter
vanishes.

\begin{figure}
\begin{center}
\includegraphics[width=3.2in,clip]
{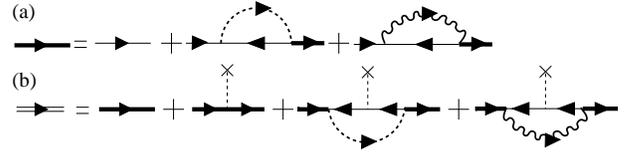}
%{scat_2G.eps}
\caption{\label{fig:scatterG} Full Green's function in (a) the clean
case (thick solid line), and (b) in the presence of an impurity
(double thin lines) to lowest order in $U_0$, the impurity strength. A thin line represents
the non-interacting normal state Green's function
$G_0=(\omega+\xi_{\bf{k}})^{-1}$, and the dotted and wavy lines
represent the $t$-matrices $t_{sc}$ and $t_{pg}$ respectively. Panels (a) and (b)
are the basis for our ARPES and QPI analysis}. 
\end{center}
\end{figure}

Our preformed pair theory \cite{ourarpespapers,ourreview},
which has also been successfully applied to the ultracold atoms,
is based on
BCS-Bose Einstein condensation (BEC) crossover theory, which focuses
on a stronger than BCS attractive interaction and the associated
short coherence length. The ground state is the usual BCS state
with a self consistently determined chemical potential. As outlined
elsewhere \cite{ourreview}, we solve
coupled equations for the fermionic and pair propagators.
In
the pseudo-gap region below $T^*$, there are preformed pairs that
become non-condensed pair excitations of the condensate below $T_c$.
For $T< T_c$, both non-condensed and condensed pairs co-exist, but
there is a \textit{gradual} conversion from
non-condensed to condensed pairs as temperature is decreased,
leading to the conventional BCS ground state as
$T \rightarrow 0$.

The pair
propagator is related to
the $t$-matrix, $t(Q)=t_{sc}(Q) +t_{pg}(Q) $ (with $Q$
defined as a four-vector), where $t_{sc}(Q) \equiv -\frac{\Delta_{sc}^2}{T}
\delta(Q)$ and $t_{pg}(Q)$ represent the contribution to the $t$-matrix from condensed and non-condensed
pairs, respectively. The non-condensed pair contribution $t_{pg}(Q)$ is obtained from a
particle-particle $t$-matrix \cite{ourarpespapers} that includes one
dressed and one bare Green's function \cite{Kadanoff}. The contribution of the
non-condensed pairs to the full self energy $\Sigma(K) \equiv
\sum_{Q} t(Q) G_0(Q-K)\varphi_{\bf k -q/2}^2$ can be well
approximated in terms of a pseudogap parameter
$\Delta_{pg}(\bf{k})$ because $t_{pg}(Q)$ is strongly peaked at small $Q$ for $T$ slightly above $T_c$ and $T\leq T_c$. Thus, $\Sigma(\bf{k},\omega)$
consists \cite{Chen4} of two terms
\begin{eqnarray}\label{eq:sigma} % \nonumber to remove numbering (before each equation)
  \Sigma({\bf k},\omega) &=&
\Sigma_{sc}({\bf k},\omega)+ \Sigma_{pg}({\bf k},\omega) \\
  \nonumber &=&
\frac{\Delta_{sc}^2({\bf k})}{\omega+\xi_{\bf k}}+\bigg[\frac{\Delta^2_{pg}({\bf k})}{\omega+\xi_{\bf k}+i\gamma}
%- i \Sigma_0
\bigg]
\end{eqnarray}
The two gap parameters, $\Delta_{sc}({\bf
k})=\Delta_{sc}\phik$ and $\Delta_{pg}({\bf k})=\Delta_{pg}\phik$,
correspond to
the superconducting (sc) order parameter and the
finite momentum pair gap (pg). The factor $\phik=[\cos(k_x)-\cos(k_y)]/2$ ensures d-wave symmetry. 
The effective gap which appears in the Bogoliubov quasi-
particle dispersion is 
$\Delta({\bf
k})=\Delta\phik$ 
with
$\Delta(T)\equiv\sqrt{\Delta_{sc}^{2}(T)+\Delta^{2}_{pg}(T)}$. 

The damping $\gamma$ distinguishes the non-condensed pairs from the condensate,
as was addressed microscopically in detail
in earlier work \cite{Malypapers}. We define the ordered phase through
the non-vanishing
superfluid density, $\rho_s$. Microscopic calculations
\cite{Chen2}
based on Ward identities
have established
that $\Sigma_{pg}$ effectively cancels in $\rho_s$ so that
$\rho_s \propto \Delta_{sc}^2$ as expected.

The Green's function is
%as in
%Fig.~\ref{fig:scatterG},
$G^{-1}({\bf k},\omega)= \omega-\xi_{\bf k} -\Sigma({\bf
k},\omega)$ [see Fig.~\ref{fig:scatterG}(a)]. Here,
$\xi_{\bf{k}}=-2t(\cos{k_x}+\cos{k_y})-4t^{\prime}\cos{k_x}\cos{k_y}-\mu$
is the normal state tight binding dispersion with $t=300$ meV,
$t^{\prime}/t=-0.4$, and $\mu(T=0)/t=-1.083$. Here, for definiteness,
we take
$T_c = 5.9~meV$ and $\Delta(T_c) = 52~meV$, with $ n = 0.85$,
representing a
moderately underdoped system ~\cite{arpesstanford}.
We
present the simplest approximation \cite{ourarpespapers} of our
microscopic theory in order to make our
calculations more accessible. To reasonable accuracy \cite{ourreview},
$\Delta(T)$ 
can be found from the BCS gap equation; we approximate \cite{ourarpespapers} 
$\Delta_{pg}(T)=\Delta(T)(T/T_c)^{3/4}$ for $T \le T_c$ and
$\Delta_{pg}(T)=\Delta(T)$ for $T>T_c$.

The link between
ARPES experiments and STM experiments
both {\it below} and {\it above} $T_c$ is based on
the common
spectral function $A({\bf
k},\omega)=-{\rm Im} G({\bf k},\omega)/\pi$.
These ARPES experiments constrain the single adjusted
parameter $\gamma$ that, nevertheless, has a well understood microscopic
origin \cite{Malypapers}. 
We find the presence of a perceptible
Fermi arc (with length $\gtrsim 10 \%$ of the Fermi surface)
requires \cite{ourarpespapers,Chubukov2}
that $\gamma(T_c)/\Delta(T_c) > 0.2$. We have importantly determined
that the behavior of QPI that we report in this Letter is robust for
the same regime,
$\gamma(T_c)/\Delta(T_c)
> 0.2$.
Specifically, to account for a Fermi arc of
about 20\% of the length of the Fermi
surface we take $\gamma
(T_c) \approx 0.5 \Delta_{pg}(T_c)$.
The behavior of the transport lifetime
(see Ref.~\cite{Chen4} and references therein)
suggests that $\gamma$ has a characteristic cubic $T$ dependence below $T_c$
and linear above: $\gamma(T)=\gamma(T_c)(T/T_c)^{3}$ for $T\le T_c$,
$\gamma(T)=\gamma(T_c)(T/T_c)$ for $T> T_c$. 
Our results are not
particularly sensitive to the detailed $T$-dependence,
which, for the present purposes, could have been ignored.
%, nor to
%$\Sigma_{0}(T)$ which we took, for definiteness, to be
%$0.1\gamma(T)$.

In the presence of a non-magnetic impurity, the diagrams
contributing to the full Green's function up to first order in the
impurity strength $U_0$ are shown in Fig.~\ref{fig:scatterG}(b).
For a single point-like impurity, the Fourier transform of the first
order correction to the LDOS is given by
\begin{eqnarray}\label{eq:fdos}
\delta n({\bf q},\omega)&=&-\frac{U_0}{\pi} {\rm Im} \left[ \int
\frac{d^2k}{(2\pi)^2} \left( G({\bf k},\omega)G({\bf k}+{\bf
q},\omega) \right. \right. \nonumber
\\
&& \hspace{-1.5cm} \left. \left. -F_{sc}({\bf k},\omega)F_{sc}({\bf
k}+{\bf q},\omega) -F_{pg}({\bf k},\omega)F_{pg}({\bf k}+{\bf
q},\omega) \right)
\right] \nonumber \\
\end{eqnarray} where
$$
F_{sc}({\bf K}) \equiv -\frac{\Delta_{sc}({\bf k}) G({\bf
K})}{\omega+\xi_{{\bf k}}}; \, F_{pg}({\bf
K}) \equiv-\frac{\Delta_{pg}({\bf k}) G({\bf K})}{\omega+\xi_{\bf k}
+i\gamma}
$$
%CHECK: (2\pi)^2
and ${\bf K}=({\bf k},\omega)$. There are two types of $FF$ terms:
the usual one which depends on $F_{sc}$ associated with the 
sc  
condensate \cite{Capriotti03}, and a new term (involving $F_{pg}$)
represented by the last diagram on the right hand side of
Fig.~\ref{fig:scatterG}(b); this reflects the contribution of
pre-formed (or non-condensed) pairs. Importantly, this term (which
has not appeared in previous work) is required for microscopic
consistency.
%Indeed, if we take $\Sigma_0 = \gamma$,
%then above $T_c$ this theory precisely reduces to the usual BCS form
%for QPI but with an overall broadening.

\begin{figure} \begin{center}
\includegraphics[width=2.5in,clip]
{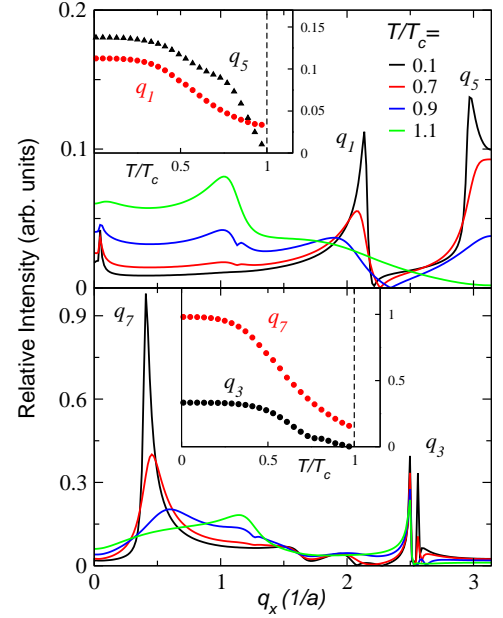}
%{cutom-10.eps}
\caption{\label{fig:tempdependence} (Color online) One dimensional
cuts 
for fixed frequency
$\omega=-10meV$ in the (a) horizontal and (b) diagonal
directions. Intensities of peaks fall off rapidly as Cooper pairs
lose coherence. Insets specify temperature dependence of $q$-vector
intensity. All octet peaks disappear at and above $T_c$, although
there is a finite background for $q_1$ and $q_7$.}
\end{center}
\end{figure}

\begin{figure*} \includegraphics[height=3.5in,clip]
{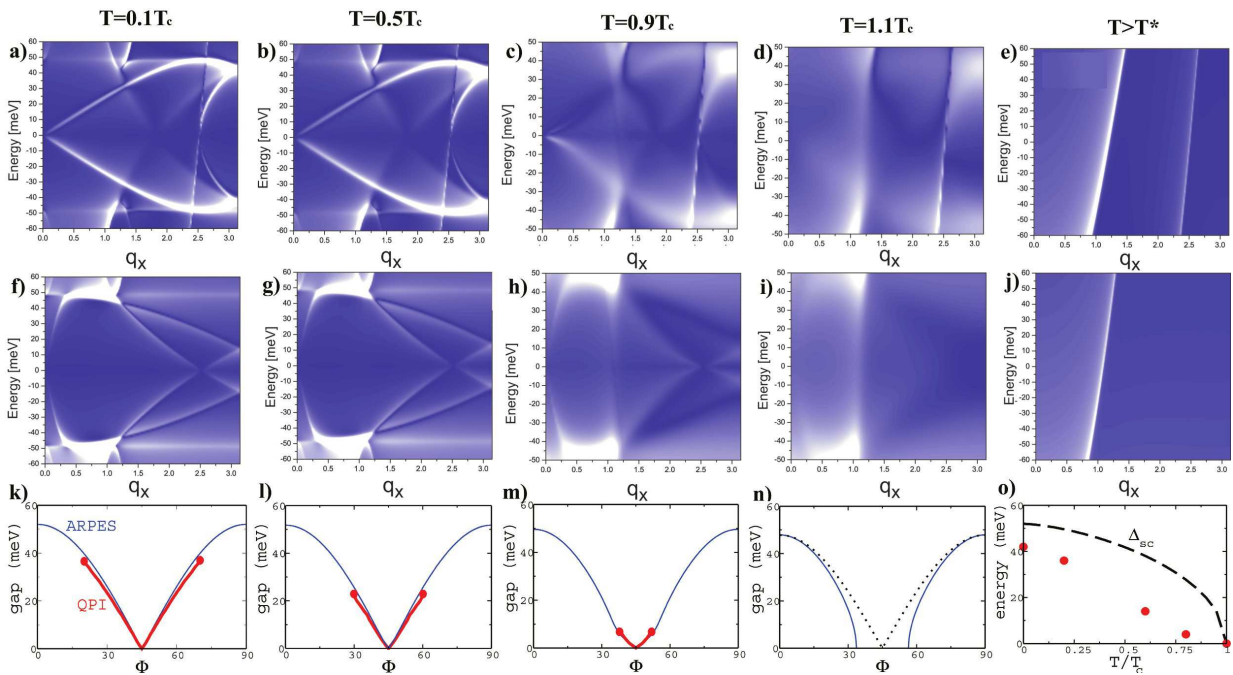}
%{fig3norings2.eps}
\caption{\label{fig:bigpanel} (Color online) 
First (Second) row shows 
diagonal (horizontal) cuts along $q_x=q_y$
($q_y=0$) at $\omega = -10$ meV
for increasing $T$ from left to
right $T/T_c = 0.1,~0.5,~0.9$ and $1.1$. 
The last row
plots
the B-QPI gap (thick red lines) at the same temperatures
inferred from pairs of octet vectors along the horizontal and diagonal directions, compared with the ARPES gap (thin blue lines).
When the
inconsistency between the inferred points on the Fermi surface, $k_x$ and $k_y$,
 exceeds $5\%$ the inversion procedure is terminated.  The last figure (3o)
on the right shows the temperature dependence of the energy where the octet model breaks down (red dots), which 
falls 
with temperature, like the order parameter (black dashed line). The dotted line in the adjacent
figure (3n) represents the simple $d$-wave gap shape. 
It is clear that, as in experiment \cite{ShenNature}
the ARPES gap varies smoothly across $T_c$ as in a second
order phase transition.
}
\end{figure*}

We have studied the evolution of the QPI pattern with temperature
and find that the
QPI pattern remains
relatively unchanged up to $T=0.5T_c$. This low $T$ behavior is similar
to that observed in previous \cite{DHLEE2} $T =0 $ calculations. However,
the peak intensity begins to
drop rapidly above this temperature.
At $T/T_c = 0.9$ the
characteristic (superconducting) QPI pattern becomes smeared out
because of the increase in the number of non-condensed pairs. The
octet peaks vanish above $T_c$.
We find that the QPI pattern above $T_c$
is also distinctively different from that of a gapless normal
phase \cite{YazdaniPRB,Capriotti03}, as expected in the 
presence of a pseudogap. 
%The presence of non-condensed
%pairs below $T_c$ and pre-formed pairs above $T_c$ guarantees,
%that the QPI pattern evolves smoothly (as in a second
%order phase transition) through $T_c$. 

In order to elucidate the temperature dependence of the QPI pattern
in more detail, we plot in Fig.~\ref{fig:tempdependence} $|\delta
n({\bf q},\omega)|$ as a function of $q$ along
the horizontal and diagonal directions for $\omega=-10meV$. 
The insets of Fig.~\ref{fig:tempdependence}(a) and (b) show the
temperature dependence of the peak intensity for the dispersing
vectors. In agreement with the above discussion, we find that the
peak intensities decrease significantly for $T>0.5 T_c$. At $T
\approx T_c$, the peaks associated with the dispersing branches
$q_1, q_3,q_5$, and $q_7$ all vanish (the peaks associated with $q_1$
and $q_7$ merge into the background and thus should not be identified as
distinctive individual peaks).
%The dispersing peaks thus act as
%indicators of superconducting coherence.
%In contrast to the dispersing features, the
%intensity of the nearly non-dispersive branches decreases only
%weakly through $T_c$ and persists up to higher temperatures since it
%is present in the $\Delta=0$ normal state \cite{Capriotti03}.
These insets show that 
the B-QPI pattern is uniquely associated with
superconducting coherence;  this strong signature of $T_c$
in QPI is correlated with its counterpart in ARPES \cite{ShenNature,Kanigel} 
for moderately
underdoped systems.

We turn now to detailed plots of the QPI peak dispersion that
reinforce these conclusions.  
Our main finding is that the
dispersive behavior vanishes in the normal state, as is
consistent with experiment \cite{Yazdani}. This is shown in
Figs.~\ref{fig:bigpanel}(a)-(e) and
Figs.~\ref{fig:bigpanel}(f)-(j)
where 
$|\delta n({\bf q},\omega)|$ is plotted as a function of $\omega$ and momentum
along the diagonal direction, $q_x=q_y$, and horizontal direction, $q_y=0$. 
We identify several non-dispersive features in the B-QPI pattern
that reflect the normal state.
It is not clear to what extent the specifics of
the non-dispersive features seen experimentally \cite{Yazdani}
above $T_c$ are present here.
In the diagonal cut at $T=0.1T_c$ and at low
frequencies and momentum $q \approx 0$, we identify the
the octet vector $q_7$ [see
Fig.~\ref{fig:bigpanel}(a)], while the peak at momentum $q \approx
2.5$ corresponds to $q_3$.
In the horizontal cut, we identify the dispersing $q_1$ and
$q_5$-branches, which for $\omega \rightarrow 0$ merge into a single
point, $q \approx 2.55$.
%The momenta $q_1, q_3, q_3$, and $q_7$ are consistently related by
%the octet model (within 1-2 percent) for sufficiently low
%frequencies.
% up to $40$ meV
%$\approx 0.8 \Delta$; a consistency which, however, breaks down for
%larger frequencies. With increasing temperature, we find that the
%intensity of the peaks in the QPI pattern decreases for $T \gtrsim
%T_c/2$, and that the entire pattern become smeared out; it evolves
%smoothly into that of the pseudo-gap state.
The momenta $q_1,q_3,q_5$ and $q_7$ satisfy the constraints of
the octet model for sufficiently low $\omega$.

In the last row of Fig. 3, we probe the effects of superconducting
coherence in another way that is strictly confined to the
$ T \leq T_c$ state. Here we display the gap obtained for
a given temperature from the B-QPI pattern dispersion shown
in the first two rows, using the octet model inversion procedure
\cite{Seamus}. Each row was used to generate two sets of points.
For definiteness, we take a $ 5 \%$
consistency level as a necessary condition for the observation of
B-QPI.
For a 
given energy $\omega$, the inversion procedure yields points on the Fermi 
surface, $k_x$ and $k_y$ such that $\Delta(k_x,k_y)=\omega$. 
The ARPES-derived gap \cite{ourarpespapers} 
is plotted for comparison. 
For frequencies $\omega >
0.8\Delta_{sc}$, we find
%the octet-derived gap cannot be reliably extracted,
%since here
the two sets of Fermi surface points extracted from the octet
vectors differ by more
than $5\%$. This leads to a frequency cut-off in the inversion
procedure which is plotted in the right-most
panel of the bottom
row as a
function of temperature, where it tracks the
superconducting order parameter $\Delta_{sc}$.
%, so that
%The inversion procedure adequately predicts a gap along the
%Fermi surface only for frequencies $\omega<\Delta_{sc}$.
%its
%size essentially
%constrains the segment of the Fermi
%surface over which the octet model is applicable.
Consistency between ARPES and B-QPI is found for all $T \leq T_c$ except
in the
normal state plot in the last row
where the Fermi arc is clearly seen in the ARPES derived spectral gap.
One can see from the panel at $T/T_c = 0.9$ that Fermi arc physics
is also apparent below $T_c$ in the distortion of the shape of the QPI-inferred gap.
Both gaps
extrapolate at the antinodes
to approximately $50 meV \approx \Delta(T)$.
The segments of ${\bf k}$ space where B-QPI inversion is possible
correspond to the
so-called \cite{Seamus} ``Bogoliubov arcs".

While our results for $T \ll T_c$ are in agreement with those of
earlier studies \cite{DHLEE2,Capriotti03}, our results for
temperatures above $T_c$ differ in several important respects from
previous work \cite{FranzQPI,YazdaniPRB}. In contrast to the
pseudogap theory proposed in Ref.~\cite{FranzQPI}, we incorporate
the presence of Fermi arcs in ARPES above $T_c$. This may explain why the
pseudogap QPI pattern of Ref.~\cite{FranzQPI} is qualitatively
similar to that of the sc state. On the other hand, in
Ref.~\cite{YazdaniPRB} the Fermi arcs were incorporated,
but under the presumption that the gap parameters entering the
calculation of the QPI pattern and the spectral gaps observed in
ARPES experiments were equivalent. The resultant QPI pattern
disagreed with the experimental results and it was therefore
suggested that the pseudogap phase may be unrelated to precursor
superconductivity. In contrast, we argue in our preformed pair
scenario that, while the same formalism must be used to address ARPES
and STM experiments (as in Fig.~\ref{fig:scatterG}), the ARPES
spectral gap should not be directly employed in the calculations of
the QPI pattern.

In summary, in our QPI studies two distinct scales appear, 
$\Delta_{sc}({\bf
k})$
and
$\Delta({\bf
k})$, which 
have the same ${\bf k}$ dependence. It 
appears naturally in these STM studies
that $\Delta_{sc}(T)$, which vanishes at $T_c$, 
is more influential in the nodal regime, as is widely inferred to be
the case \cite{Seamus,ShenNature}.
By contrast $\Delta$ is roughly $T$ independent.
We extract $\Delta_{sc}(T)$ from the
``${\bf k}$-space extinction" of the QPI-inferred gap.
(As in Reference \cite{Shen08}, we do not find this
${\bf k}$-space- extinction point
to be directly related to the
magnetic zone boundary, as claimed elsewhere \cite{Seamus}.)
We extract $\Delta$ from the Bogoliubov quasi-particle
dispersion, as obtained from the B-QPI inversion procedure.
A difference with Reference \cite{Seamus} is that we find
the inversion procedure works well even down to the precise nodal
point. The failure of this procedure when experimentally
implemented near the nodes has not
been understood.

An important feature of the present theory is its
unique ability to address the ordered phase at general $T$ in the presence of
a pseudogap. Other theories are confined to
$ T \approx 0$ or $ T > T_c$.
Because of this pseudogap it is difficult, but important
for theorists to find ways to help identify the very fundamental
superconducting order parameter, as we do here.
While our microscopically based theory \cite{ourreview} contains one
adjusted parameter ($\gamma(T_c)$) our theory should not be
viewed as principally 
phenomenological. 
%We can take $\gamma$ to be
%$T$ independent for the present purposes. Additionally, 
We have found an important
correlation which makes our predictions
very robust. The condition on $\gamma(T_c)$
for the appearance of perceptible Fermi arcs
in photoemission
%($\gamma(T_c) /\Delta(T_c) > 0.2$)
is
essentially the same as the condition for the disappearance of
octet peaks in the normal state.
In heavily
underdoped cuprates (not considered here) we cannot rule out a small
contribution of B-QPI in the near vicinity but above $T_c$
presumably reflecting the fact that in ARPES
the Fermi arcs are not present \cite{Shen08} until substantially above
the nominal $T_c$. For these more metallurgically
complex cuprates near the insulating phase,
more detailed experiments and theory
are needed to probe the
correlations between B-QPI and ARPES.

\textit{Note added.} After this work was submitted we learned of related STM experiments 
but on heavily
underdoped cuprates by Pushp \textit{et al.} \cite{Yazdani3}.

This work was supported by Grant Nos. NSF PHY-0555325 and NSF-MRSEC
DMR-0213745 and by the U.S. Department of Energy under Award No.
DE-FG02-05ER46225 (D.M.). We thank J.C. Davis, J. Lee and A. Yazdani
for helpful discussions.

\bibliographystyle{apsrev} 
%\bibliography{Review2} 

\end{document}